      [1999/12/01 v1.4c Il Nuovo Cimento]
\documentclass[preprint]{cimento}

\usepackage{cite}
\usepackage{cancel}
\usepackage{amsmath}
\usepackage{graphicx}
\newcommand{\eq}[1]{Eq.~(\ref{#1})}
\newcommand{\ps}{p\hspace{-0.44em}/\hspace{0.06em}}
\newcommand{\Msusy}{M_{\textrm{SUSY}}}

\newcommand{\gev}{\, \mathrm{GeV}}
\newcommand{\tev}{\, \mathrm{TeV}}

\title{New Physics in Flavour Observables}
\author{A.~Crivellin\from{ins:x}}
\instlist{\inst{ins:x} Paul Scherrer Institut, CH--5232 Villigen PSI, Switzerland}
\PACSes{11.30.Hv,12.60.Cn,12.60.Fr,13.25.Hw,14.65.Fy,14.80.Da,14.80.Ly}
\begin{document}
\maketitle
\begin{abstract}
LHCb found hints for physics beyond the Standard Model (SM) in $B\to K^*\mu^+\mu^-$, $R(K)$ and $B_s\to\phi\mu^+\mu^-$. These intriguing hints for NP have recently been confirmed by the LHCb measurement of $R(K^*)$ giving a combined significance for NP above the $5\,\sigma$ level. In addition, the BABAR, BELLE and LHCb results for $B\to D^{(*)}\tau\nu$ also point towards lepton flavour universality (LFU) violating new physics (NP). Furthermore, there is the long-standing discrepancy between the measurement and the theory prediction of the anomalous magnetic moment of the muon ($a_\mu$) at the $3\,\sigma$ level.
Concerning NP effects, $b\to s\mu^+\mu^-$ data can be naturally explained with a new neutral gauge bosons, i.e. a $Z^\prime$ but also with heavy new scalars and fermions contributing via box diagrams. Another promising solution to $b\to s\mu^+\mu^-$, which can also explain $B\to D^{(*)}\tau\nu$, are leptoquarks. Interestingly, leptoquarks provide also a viable explanation of $a_\mu$ which can be tested via correlated effects in $Z\to\mu^+\mu^-$ at future colliders. Considering leptoquark models, we show that an explanation of $B\to D^{(*)}\tau\nu$ predicts an enhancement of $b\to s\tau^+\tau^-$ processes by around three orders of magnitude compared to the SM. In case of a simultaneous explanation of $B\to D^{(*)}\tau\nu$ and $b\to s\mu^+\mu^-$ data, sizable effects in $b\to s\tau\mu$ processes are predicted.
\end{abstract}

\section{Introduction}

With the discovery of the Brout--Englert--Higgs boson the LHC completed the SM of particle physics and the main focus shifted towards the discovery of new particles and new laws of physics. While no new particles have been discovered directly at the LHC so far, some intriguing 'hints' for indirect effects of NP in the flavor sector appeared. In $b\to s\ell^+\ell^-$ transitions and in $B\to D^{(*)}\tau\nu$ significant deviations from the SM predictions appeared. In addition, there is the long standing anomaly in the anomalous magnetic moment of the muon. In these proceedings we review these anomalies and how they can be explained within some selected NP models.

\section{$b\to s\ell^+\ell^-$}

LHCb reported deviations from the SM predictions in the angular observable called $P_5^\prime$~\cite{Descotes-Genon:2013vna} in $B\to K^* \mu^+\mu^-$ and in the decay $B_s\to\phi\mu^+\mu^-$~\cite{Aaij:2013qta,Aaij:2013aln} with a significance of $2$--$3\,\sigma$ each. Furthermore, LHCb~\cite{Aaij:2014ora} found indications for the violation of lepton flavour universality in $R(K)={\rm Br}[B\to K \mu^+\mu^-]/{{\rm Br}[B\to K e^+e^-]}=0.745^{+0.090}_{-0.074}\pm 0.036\,,$ in the range $1\,{\rm GeV^2}<q^2<6\,{\rm GeV^2}$ which disagrees with the theoretically clean SM prediction $R_K^{\rm SM}=1.0003 \pm 0.0001$~\cite{Bobeth:2007dw} by $2.6\,\sigma$. Combining these anomalies with all other observables for $b\to s \mu^+\mu^-$ transitions, it is found that a scenario with NP in $C_9^{\mu\mu}$ only, is preferred compared to the SM by more than $4\,\sigma$ \cite{Altmannshofer:2015sma,Descotes-Genon:2015uva,Hurth:2016fbr}. Recently, LHCb confirmed these hints for NP by measuring $R_{K^*}$~\cite{Aaij:2017vbb}. Now, a combined analysis of these observables gives a significance for NP above the $5\,\sigma$ level~\cite{Capdevila:2017bsm}.

A rather large contribution to the operator $(\overline{s}\gamma_\alpha P_L b)(\overline{\mu}\gamma^\alpha \mu)$, as required by the model independent fit, can be achieved in models containing a heavy $Z^\prime$ gauge boson (see for example~\cite{Gauld:2013qba,Buras:2013dea} for some early analysis). If one aims at explaining also $R(K)$ and $R(K^*)$, a contributing to $C_9^{\mu\mu}$ involving muons is necessary but to $C_9^{ee}$ with electrons no contribution is required. This is naturally the case in models with gauged muon minus tau number ($L_\mu-L_\tau$)~\cite{Altmannshofer:2014cfa,Crivellin:2015mga,Crivellin:2015lwa,Crivellin:2016ejn}. In these $Z^\prime$ models unavoidable contributions to $B_s--\overline{B}_s$ are generated which constrain the coupling to muons to be much larger than the one to $\bar s b$ unless there is a fine-tuned cancellation~\cite{Crivellin:2015era}.

An alternative solution involves leptoquarks. Here, the only (single) representation of scalar leptoquarks which gives a good fit to data via a tree-level contribution in the from of a $C_9=-C_{10}$ solution is the $SU(2)$ triplet. 

Since in $b\to s\ell^+\ell^-$ processes one competes with a loop and CKM suppressed SM contribution, also a NP model with a loop effect is capable of explaining the data. In fact, it has been shown in Ref.~\cite{Gripaios:2015gra} and~\cite{Arnan:2016cpy} the boxes with heavy new scalars and fermions can achieve this. Here, one needs to add at least three particles to the SM content: One additional fermion $\Psi$ and two additional scalars $\Phi_Q$ and $\Phi_\ell$ or two additional fermions $\Psi_Q$ and $\Psi_\ell$ and one additional scalar $\Phi$. In both cases the additional particles interact with left-handed $b$-quaks, $s$-quarks and muons via Yukawa-like couplings $\Gamma_b$, $\Gamma_s$ and $\Gamma_\mu$, respectively one finds a $C_9=-C_{10}$ like contribution to $b\to s\mu^+\mu^-$.

We consider all representations under $SU(2)$ and $SU(3)$ which are also present in the SM. Here we have the following possibilities~\cite{Arnan:2016cpy}:
\begin{equation}
\begin{array}{*{20}{c}}
\begin{array}{*{20}{c}}
{SU\left( 2 \right)}&\vline& {{\Phi _Q},{\Psi _Q}}&{{\Phi_\ell},{\Psi_\ell}}&{\Psi ,\Phi }\\
\hline
I&\vline& 2&2&1\\
II&\vline& 1&1&2\\
III&\vline& 3&3&2\\
IV&\vline& 2&2&3\\
V&\vline& 3&1&2\\
VI&\vline& 1&3&2\\
\end{array}\quad\quad\quad\quad
\begin{array}{*{20}{c}}
{SU\left( 3 \right)}&\vline& {{\Phi _Q},{\Psi _Q}}&{{\Phi_\ell},{\Psi_\ell}}&{\Psi ,\Phi }\\
\hline
A&\vline& 3&1&1\\
B&\vline& 1&{\bar 3}&3\\
C&\vline& 3&8&8\\
D&\vline& 8&{\bar 3}&3\\
\end{array}
\end{array}
\label{eq:reps}
\end{equation}

\begin{figure}[t]
\begin{center}
\begin{tabular}{cp{7mm}c}
\includegraphics[width=0.6\textwidth]{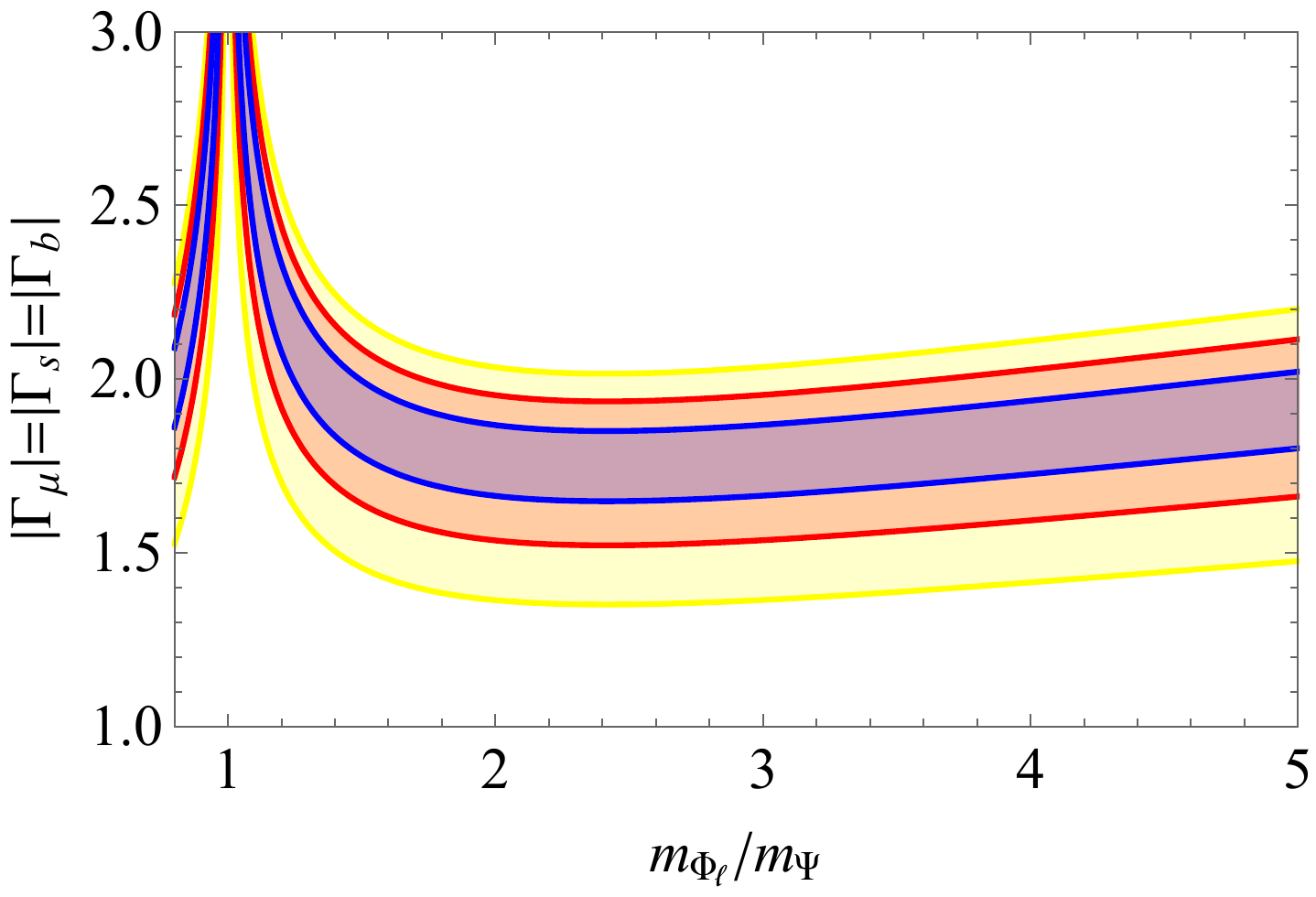}
\end{tabular}
\end{center}
\caption{Allowed regions for the coupling strength to muon, bottom and strange quarks from $b\to s\mu^+\mu^-$ data as a function of $m_{\Phi_\ell}/m_\Psi$ for case A-I in the scenario with $m_{\Phi_Q}=m_\Psi=1\,$TeV. Blue, red and yellow correspond to $1\,\sigma$, $2\,\sigma$ and $3\,\sigma$, respectively.
 \label{GammaVSxL} }
\end{figure}

The most stringent bounds on this model come from $B_s--\overline{B}_s$ mixing. In fact, it turns out that in order to get a sizable effect in $b\to s\mu^+\mu^-$ one needs a cancellation in $B_s--\overline{B}_s$ mixing. This is (for real couplings) only possible for Majorana representations for $\Psi$ in which crossed and uncrossed box diagrams enter with opposite sign, i.e. for the representations A-I, A-IV, C-I and C-IV with vanishing hypercharge. For A-I the coupling strength necessary to explain $b\to s\mu^+\mu^-$ is shown in Fig.~\ref{GammaVSxL}.

\section{$R(D)$ and $R(D*)$}

Hints for LFU violating NP also comes from the semileptonic $B$ decays $B\to D^{(*)}\tau\nu$ where the combination of the measured ratios exceed the SM ones by $3.9\,\sigma$~\cite{Amhis:2016xyh}.

An explanation of $R(D)$ and $R(D^*)$ is not easy. Since these processes are mediated in the SM already at tree-level, a rather large NP contribution is required to account for the $O(20\%)$ deviation. Therefore, new particles added to the SM for explaining $R(D)$ and $R(D^*)$ cannot be very heavy and must have sizable couplings. In the past, mainly three kinds of models with the following new particles have been proposed:
Charged Higgses (e.g.~\cite{Crivellin:2012ye,Crivellin:2013wna}), $W^\prime$ gauge bosons (e.g.~\cite{Greljo:2015mma}) and leptoquarks (e.g.~\cite{Fajfer:2012jt,Alonso:2015sja,Calibbi:2015kma,Barbieri:2015yvd}). 

Models with charged Higgses lead to (too) large effects in the total $B_c$ lifetime~\cite{Alonso:2016oyd} and, depending on the coupling structure, can also be disfavored by the $q^2$ distribution~\cite{Freytsis:2015qca,Celis:2016azn}. Interestingly, if the couplings of the charged Higgs are chosen in such a way that they are compatible with the measured $q^2$ distribution, these models are ruled out by direct searches~\cite{Faroughy:2016osc}.
Models with $W^\prime$ gauge bosons are also delicate because they necessarily involve $Z^\prime$ bosons due to $SU(2)$ gauge invariance. If the $Z^\prime$ width is not unnaturally large, these models are again ruled out by direct searches~\cite{Faroughy:2016osc}. In models with leptoquarks generating left-handed vector operators the coupling structure should be aligned to the bottom quark in order to avoid $b\to s\nu\nu$ bounds. However, in this case the effect in $R(D)$ and $R(D^*)$ is proportional to the small CKM element $V_{cb}$ and large third generation couplings are required to account for the anomalies. These large third generation couplings lead again to stringent bounds from direct LHC searches~\cite{Faroughy:2016osc} and electroweak precision observables~\cite{Feruglio:2016gvd}.

\begin{figure}[t]
\begin{center}
\begin{tabular}{cp{7mm}c}
\includegraphics[width=0.41\textwidth]{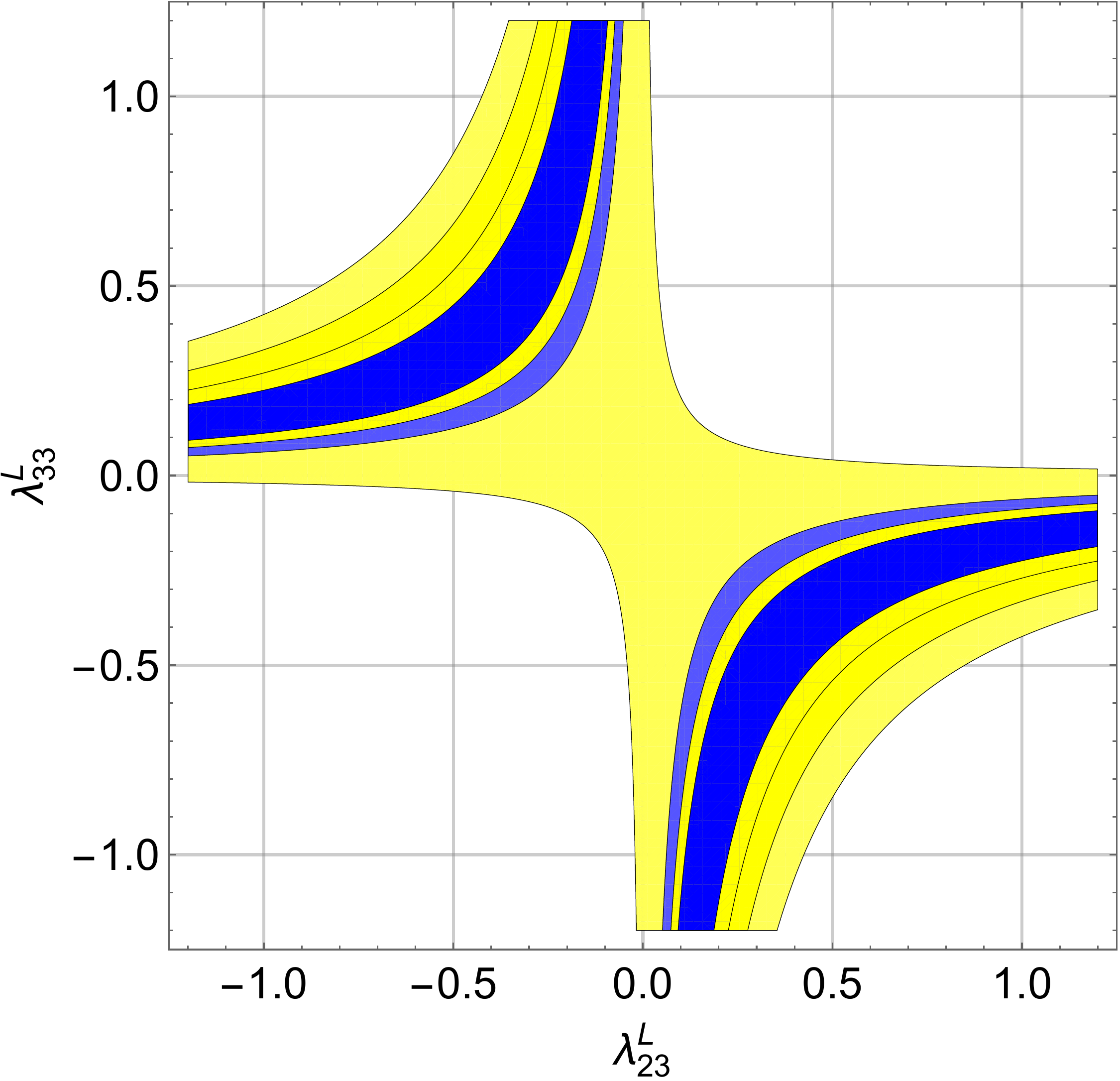}
\includegraphics[width=0.58\textwidth]{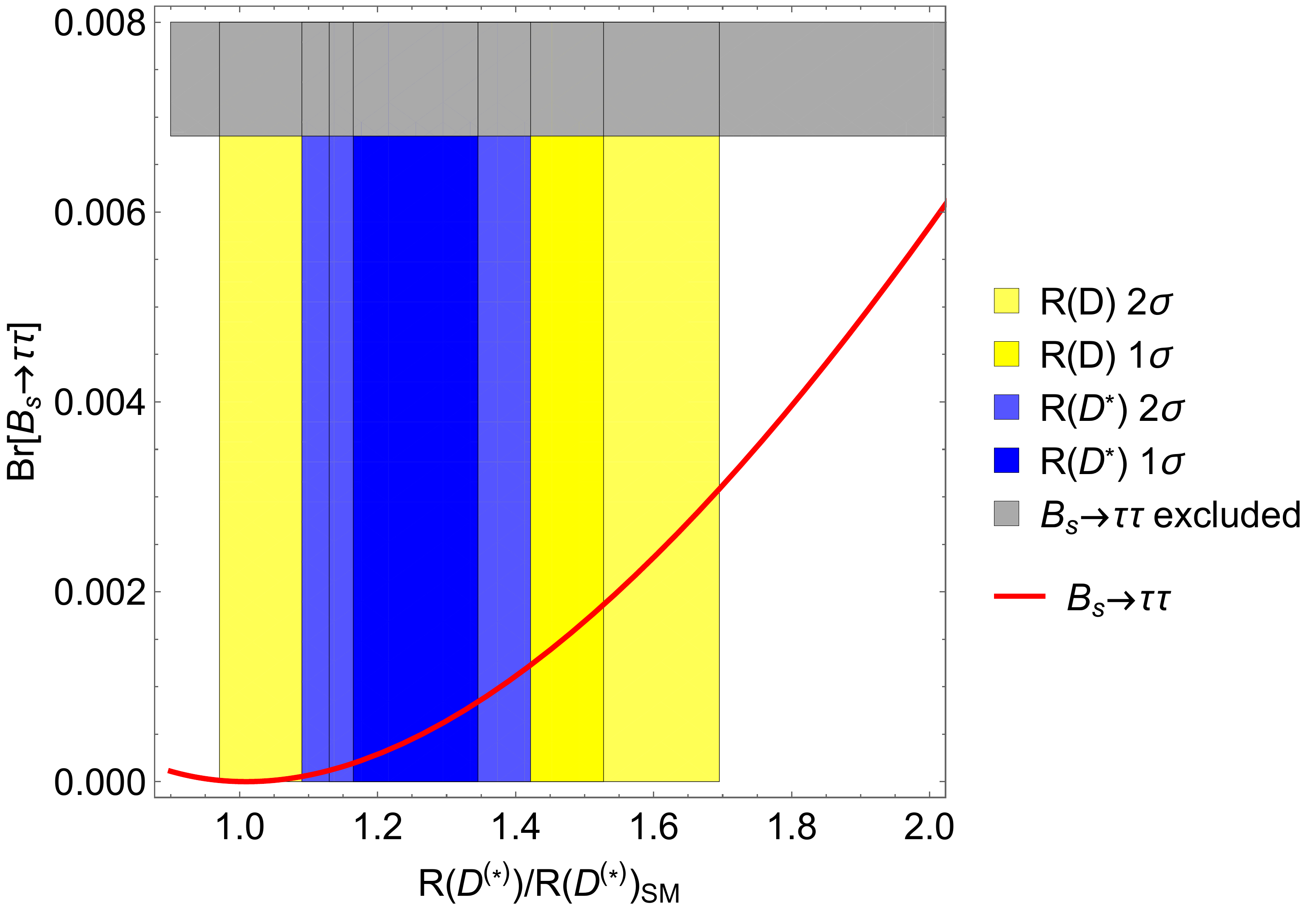}
\end{tabular}
\end{center}
\caption{Left: Allowed regions by $R(D)$ and $R(D^*)$ in the $\lambda^L_{23}-\lambda^L_{33}$ plane for a leptoquark mass of $M=1\,$TeV. Note that already small couplings are sufficient to account for $R(D)$ and $R(D^*)$. Therefore, the bounds from LHC searches are weakened and the leptoquarks can also be easily heavier than $1\,$TeV and still explain the anomalies with couplings in the perturbative regime. Right: Prediction for $B_s\to\tau\tau$ (red) as a function of $R(D^{(*)})/R(D^{(*)})_{\rm SM}$. Here we neglected small CKM suppressed contributions.}         
\label{Bstautau}
\end{figure}

In Ref.~\cite{Crivellin:2017zlb} it has been shown that two scalar leptoquarks with related couplings (or a vector leptoquark singlet) can explain $R(D)$ and $R(D^*)$ without violating bounds from direct searches or EW precision data. The key point is that no effects in $b\to s\nu\nu$ transitions are generated such that non-CKM suppressed effects in $R(D^{(*)})$ are possible. Therefore, already quite small couplings are sufficient to account for the tauonic $B$ decays (see left plot in Fig.~\ref{Bstautau}). However, large effects in $b\to s\tau^+\tau^-$ transitions are predicted which are in the reach of future LHCb searches (see right plot of Fig.~\ref{Bstautau}).

\section{Anomalous magnetic moment of the muon}

\begin{figure}[tb]
\begin{center}
\begin{tabular}{cp{7mm}c}
\includegraphics[width=0.58\textwidth]{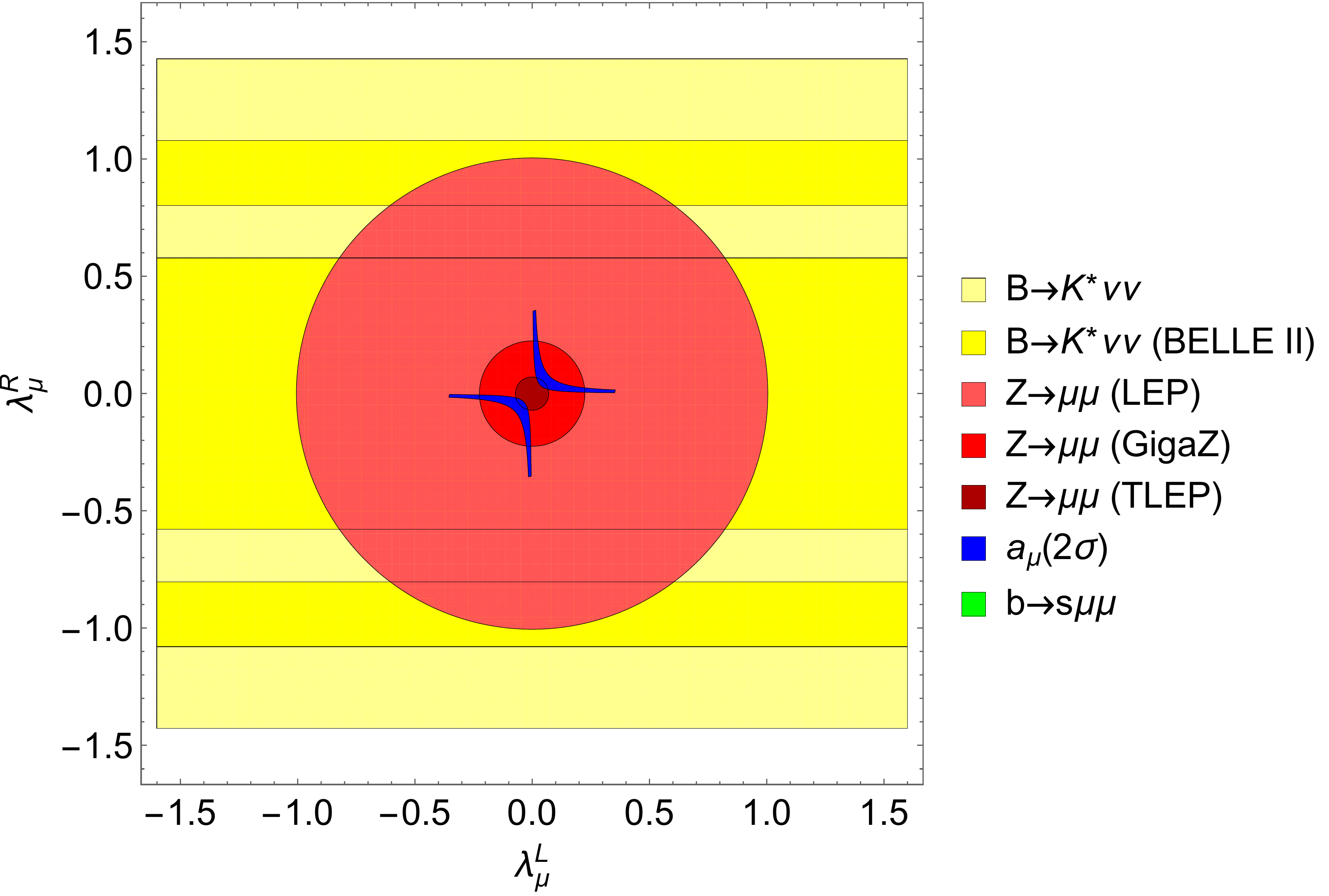}
\includegraphics[width=0.42\textwidth]{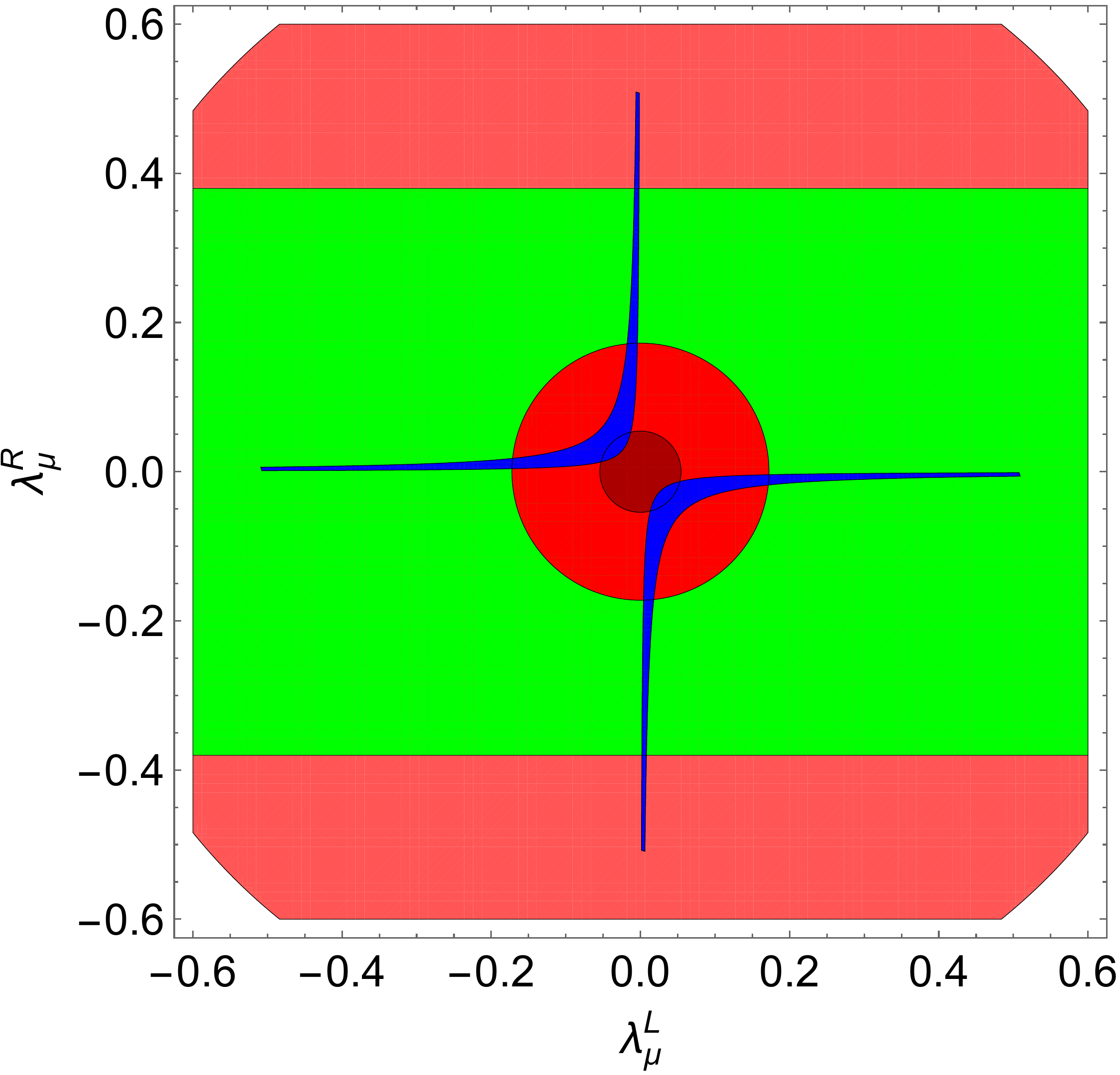}
\end{tabular}
\end{center}
\caption{Left: Allowed regions in the $\lambda^L_\mu$-$\lambda^R_\mu$ plane from current and future experiments for $SU(2)$ singlet leptoquarks $\Phi_1$ with $M=1\,$TeV. Right: Same as the left plot for the $SU(2)$ doublet leptoquark $\Phi_2$.}         
\label{muLmuR}
\end{figure}

The measurement of $a_\mu \equiv (g-2)_\mu/2$ is completely dominated by the Brookhaven experiment E821~\cite{Bennett:2006fi}
$a_\mu^\mathrm{exp} = (116\,592\,091\pm54\pm33) \times 10^{-11}$
where the first error is statistical and the second one is systematic. The current SM prediction (see for example~\cite{Nyffeler:2016gnb})
$	a_\mu^\mathrm{SM} = (116\,591\,811\pm62) \times 10^{-11}$
where almost the whole uncertainty is due to hadronic effects. This amounts to a discrepancy between the SM and the experimental value of 
$ a_\mu^\mathrm{exp}-a_\mu^\mathrm{SM} = (278\pm 88)\times 10^{-11}$ 
i.e.~a $3.1\sigma$ deviation. Possible NP explanations besides supersymmetry (see for example Ref.~\cite{Stockinger:2006zn} for a review) additional fermions~\cite{Freitas:2014pua}, new scalars~\cite{Crivellin:2015hha,Altmannshofer:2016oaq,Batell:2016ove} or leptoquarks~\cite{Djouadi:1989md,Bauer:2015knc,ColuccioLeskow:2016dox}. Here, even though the leptoquark must be rather heavy due to LHC constraints, one can still get sizable effects in the AMM since the amplitude can be enhanced by $m_t/m_\mu$ compared to the SM. In fact, among the 5 scalar leptoquark representations which are invariant under the SM gauge group~\cite{Buchmuller:1986zs}, only two can in principle generate these enhanced effects as they possess couplings to left- and right-handed muons simultaneously: $\Phi_1$being an $SU(2)_L$ singlet with hypercharge $-2/3$ and $\Phi_2$ being an $SU(2)_L$ doublet with hypercharge $-7/3$. These leptoquarks lead to simultaneous effects in $Z\to\mu^+\mu^-$ couplings which can be measured at future colliders as shown in Fig.~\ref{muLmuR}.

\section{Conclusion}

In these proceedings we reviewed the implications of the deviations from the SM predictions in $b\to s\mu^+\mu^-$, $B\to D^{(*)}\tau\nu$ and $a_\mu$ for NP models. A prime candidate for the explanation of the anomaly in $b\to s\mu^+\mu^-$ data is a $Z^\prime$ boson but also models with box contributions from new heavy scalars and fermions provide a viable solution. $B\to D^{(*)}\tau\nu$ can be most naturally explained by leptoquarks which can also account for $a_\mu$ and $b\to s\mu^+\mu^-$. An explanation of $a_\mu$ with leptoquarks predicts effects in $Z\to\mu^+\mu^-$ which are  measurable at future colliders. $B\to D^{(*)}\tau\nu$ calls for a strong enhancement of $B_s\to\tau^+\tau^-$ and a simultaneous explanation of $B\to D^{(*)}\tau\nu$ and $b\to s\mu^+\mu^-$ predicts sizable rates for $b\to s\tau\mu$ processes.

\acknowledgments

I thank the organizers, especially Gino Isidori, for the invitation to "Les Rencontres de Physique de la Valle d'Aoste" and the opportunity to present these results. This work is supported by an Ambizione Grant of the Swiss National Science Foundation (PZ00P2\_154834).

\end{document}